\documentclass[10pt,twocolumn,showpacs,aps,floatfix]{revtex4}
\usepackage{epsfig}
\usepackage{dcolumn}
\usepackage{bm}

\begin{document}



\title{Axial anomaly contribution to the parity nonconservation  effects in atoms and ions.}

\author{Gavriil Shchedrin$ ^{1}$ and Leonti Labzowsky$ ^{1,2}$}

\affiliation{$^{1}$V. A. Fock Institute of Physics, St. Petersburg State University, Uljanovskaya 1,
Petrodvorets, St. Petersburg 198904, Russia}
\affiliation{$^{2}$Petersburg Nuclear Physics Institute,
Gatchina, St. Petersburg 188350, Russia}


\begin{abstract}
The contribution of the axial triangle anomalous graph to the
parity non-conservation effect in atoms is evaluated. The final
answer looks like the emission of the electric photon by the
magnetic dipole. The relative contribution to the parity
non-conservation effect in neutral atoms appears to be negligible
but is essentially larger in case of multicharged ions.
\end{abstract}



\maketitle


The problem of testing the standard model (SM) in the low-energy
physics is one of the interesting topics in physics during the last few
decades. The SM in the low energy limit is tested in particular by observing the parity nonconservation (PNC) effects in atoms. The most accurate of these experiments is the experiment with the neutral Cs atom, first proposed in \cite{B} and performed with the utmost precision in \cite{W}.\\
The basic transition, employed in the Cs experiment was the
strongly forbidden $6s-7s$ transition with the absorption of $M1$
photon. In the real experiment this very weak transition was
opened by the external electric field but it does not matter for
our further derivations. The Feynman graphs illustrating the PNC
effect in Cs are given in Figs. 1(a) and 1(b).

The atomic experiments are indirect and require very accurate
calculations of the PNC effects in Cs to extract the value of the
free parameter of the SM, the Weinberg angle which can be compared
with the corresponding high-energy value. The main difficulty with the PNC calculations in neutral atoms is the
necessity to take into account the electron correlation within the
system of all electrons. Therefore the experiments with much
simpler systems, such as the few-electron highly charge ions (HCIs)
would be highly desirable. Several proposals on the subject were
considered in \cite{Sof1, D, Lab1, Bud}.

The radiative corrections to the PNC effect appeared to be
important in Cs calculations to reach the agreement with the high
energy SM predictions. These radiative corrections include
electron self-energy, vertex and vacuum polarization corrections.
They are even more important in the case of the HCI. The entire set
of these corrections for neutral Cs atom was calculated in \cite{Fl, Mil, Sh, flamb}. The electron self-energy and
vertex corrections for HCI were obtained in \cite{Sap}; the vacuum polarization
correction was given in \cite{Lab2}.

However, the full set of radiative corrections including $Z$-boson
loops is not yet calculated, neither for neutral
Cs nor for the HCI. Therefore the problem cannot be considered as
fully solved.

In the present work we will consider a very special radiative
correction to the PNC effect, presented by a triangle Feynman
graph, or axial anomaly (AA). We understand the triangle AA as a fermion loop
with at least one weak vertex \cite{Weinberg}. Our conclusion will be that in
neutral Cs the contribution of the axial anomaly is
negligible, but in HCI it is comparable with the electron self-energy, vertex, and vacuum polarization corrections.

The leading contribution of the AA to the atomic PNC effect is depicted
in Fig. 1.c. This contribution corresponds to the Adler-Bell-Jackiw
anomaly \cite{ABJ}. In this work we will concentrate exclusively
on this term.

We employ the standard expression for the effective parity
nonconserving interaction of the atomic electron with the nucleus
\cite{X} in the form $H_{W}=A_{PNC}\rho_{N}(\vec{r})\gamma_{5}$, with
$A_{PNC}=-G_{F}Q_{W}/2\sqrt{2}$, where $G_{F}$ is the
Fermi constant and $Q_{W}$ is the weak charge of the nucleus:\\
$Q_{W}=-N+Z(1-4\sin^{2}\theta_{w})$ where Z and N are the numbers of protons
and neutrons in the nucleus, and $\theta_{w}$ is the Weinberg angle.
The recently accepted value for this parameter deduced from all available experiments in the high and low energy physics is $\sin^{2}\theta_{w}\approx{0.23}$. The function
$\rho_{N}(\vec{r})$ represents the nucleon density distribution
within the nucleus, and the $\gamma_{5}$ is the Dirac pseudoscalar matrix.
\begin{figure}[h!]
\centerline{\mbox{\epsfxsize=10cm \epsffile{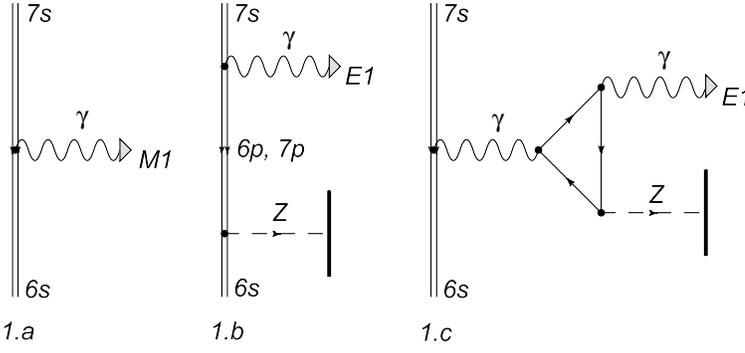}} } \caption{
The Feynman graphs that describe PNC effect in Cs. The double
solid line denotes the electron in the field of the nucleus.
The wavy line denotes the photon (real or virtual) and the
dashed horizontal line with the short fat solid line at the end
denotes the effective weak potential, i.e. the exchange by $Z$-boson
between the atomic electron and the nucleus. Graph (a)
corresponds to the basic $M1$ transition amplitude, the graph (b)
 corresponds to the $E1$ transition amplitude, induced by the
effective weak potential.
The latter violates the spatial parity
and allows for the arrival of np-states in the electron propagator
in Fig 1.b, of which the contribution of $6p, 7p$ states states
dominates. The standard PNC effect arises due to the interference between
amplitude graphs (a) and (b). Graph (c) corresponds to the axial anomaly.
The ordinary solid line represents the free electron.
To graph (c) the graph with interchanged external photon and $Z$-boson lines should be added.
}\label{helep}
\end{figure}
We write down the $S$-matrix element corresponding to the amplitude
Fig. 1(c) in the momentum representation:

\begin{widetext}
\begin{eqnarray}\label{S1}
S=-4\pi{e}^{3}\int{\frac{d^{4}p_{1}'}{(2\pi)^{4}}}
\frac{d^{4}p_{1}}{(2\pi)^{4}} \frac{d^{4}p}{(2\pi)^{4}}
\frac{d^{4}p_{2}}{(2\pi)^{4}}\hspace{1mm}
\overline{\Psi}_{n's}(p_{1})\gamma^{\rho}\Psi_{ns}(p_{1}')
\frac{g_{\rho{\nu}}}{q^{2}+i\epsilon}\nonumber\\\times
Tr\left[\gamma^{\mu}\frac{\not{p}+m_{e}}{p^{2}-m_{e}^{2}}
\gamma^{\nu}\frac{\not{p}+\not{q}+m_{e}}{(p+q)^{2}-m_{e}^{2}}
\gamma^{\lambda}\gamma^{5}\frac{\not{p}+\not{k}+m_{e}}{(p+k)^{2}-m_{e}^{2}}\right]
V^{PNC}_{\lambda}(q-k)A_{\mu}(p_{2}-k).
\end{eqnarray}
\end{widetext}

Here $e$ and $m_{e}$ are the electron charge and mass,\\
$\Psi_{ns}(p)=\Psi_{ns}(\vec{p})\delta(p_{0}-\epsilon_{ns}) $ are
the wave functions of the atomic bound electron in the state $ns$ with
$\epsilon_{ns}$ being the energy of this state; $
A_{\mu}(p_{2}-k)=(2\pi)^{4}\sqrt{2\pi/k_{0}}e_{\mu}\delta(p_{2}-k)$
is the wave function of the emitted photon, where $k_{0}\equiv{\omega}$
is the frequency and $e_{\mu}$ are 4-vector of the polarization
for this photon; and $g_{\mu{\nu}}$ is the pseudo-Euclidean metric
tensor.

The potential $V^{PNC}_{\lambda}$ for the parity nonconserving
interaction of the electron with the nucleus looks like $
V^{PNC}_{\lambda}(q-k)=A_{PNC}(q-k)\delta_{\lambda0} $, where
$\rho(q-k)=(2\pi)\delta(q_{0}-k_{0})$, $q=p_{1}-p_{1}'$ is the
transferred momentum and $k$ is the momentum of emitted photon. In
 Eq(\ref{S1}) we use the relativistic units: $\hbar=c=1$.

To begin with we consider $Z$-boson [with spin $J(Z)=1$] decay into two photons
\cite{Anomaly}. The Landau theorem forbids this
decay because two-photon system can not exist with full momentum
$J=1$ \cite{Landau1, Ax} in contrast to the allowed decay $\pi_{0}\rightarrow{\gamma\gamma}$ since $J(\pi_{0})=0$ \cite{Zuber}. We shall see this directly from the
$S$-matrix element and see also nonzero contribution in the $S$-matrix
element corresponding to the virtual photon as in our case.

The $S$-matrix element is proportional to

\begin{widetext}
\begin{eqnarray}\label{S2}
S_{\mu\nu\lambda}(k_{1},k_{2})=\int{d^{4}p}\hspace{2mm}
Tr\left[\gamma^{\mu}\frac{\not{p}\hspace{1mm}+m_{e}}{p^{2}-m_{e}^{2}}
\gamma^{\nu}\frac{\not{p}-\not{k}_{2}+m_{e}}{(p-k_{2})^{2}-m_{e}^{2}}
\gamma^{\lambda}\gamma^{5}\frac{\not{p}+\not{k}_{1}+m_{e}}{(p+k_{1})^{2}-m_{e}^{2}}\right]
\end{eqnarray}
\end{widetext}

One has to note that Eq(\ref{S2}) turns to the integral over the the loop
in Eq(\ref{S1}) under the change of variables\\ $ k_{1}\rightarrow{k};
k_{2}\rightarrow{-q} $.
Due to the identity\\
$
Tr\left[\gamma^{5}\gamma^{\tau}\gamma^{\mu}\gamma^{\nu}\gamma^{\lambda}\right]=4i\varepsilon_{\tau\mu\nu\lambda},
$
where $\varepsilon_{\tau\mu\nu\lambda}$ is the unit antisymmetric
tensor of the IV rank with definition $\varepsilon_{0123}=-1$, we
will have nonzero contribution in Eq(\ref{S2}) if and only if
we retain one momentum (with one $\gamma$-matrix) or three momenta
(with three $\gamma$-matrices) in the square bracket in
Eq(\ref{S2}). All other combinations will give zero result.
The integrals over loop with expressions containing
three momenta are convergent and could be calculated using the standard
Feynman parametrization technique \cite{Ax}. But the integrals
with expressions containing one momentum are divergent and additional
conditions are necessary to get rid of these divergences.
These conditions consist of demanding a gauge-invariance of the $S$-matrix element and look like
\begin{equation}\label{GI1}
k_{1\mu}S_{\mu\nu\lambda}=0
\end{equation}
\begin{equation}\label{GI2}
k_{2\nu}S_{\mu\nu\lambda}=0
\end{equation}
After imposing Eqs(\ref{GI1}) and (\ref{GI2}) on the $S$-matrix element (\ref{S2}) it becomes
gauge-invariant and is presented by the finite expression (it is depicted in the Fig. 1.c with additional graph with interchanged external photon and $Z$-boson lines):

\begin{eqnarray}\label{S3}
S_{\mu\nu\lambda}(k_{1},k_{2})=
J_{110}(k_{1},k_{2})
{\varepsilon_{\mu\nu\alpha\beta}}k_{1\alpha}k_{2\beta}(k_{1}+k_{2})_{\lambda}+\\\nonumber
J_{101}(k_{1},k_{2})
({\varepsilon_{\lambda\nu\alpha\beta}}k_{1\alpha}k_{2\beta}k_{1\mu}+k^{2}_{1}{\varepsilon_{\lambda\mu\nu\alpha}}k_{2\alpha})\\\nonumber-
J_{011}(k_{1},k_{2})
({\varepsilon_{\lambda\mu\alpha\beta}}k_{1\alpha}k_{2\beta}k_{2\nu}+k^{2}_{2}{\varepsilon_{\lambda\mu\nu\alpha}}k_{1\alpha})
\end{eqnarray}

where

\begin{eqnarray}\label{Int1}
J_{rst}(k_{1},k_{2})=-\frac{1}{\pi^{2}}\int^{1}_{0}
d{\xi_{1}}\int^{1}_{0}d{\xi_{2}}\int^{1}_{0}d{\xi_{3}}\\\nonumber
\frac{({\xi_{1}}^{r}{\xi_{2}}^{s}{\xi_{3}}^{t})
\delta(1-\xi_{1}-\xi_{2}-\xi_{3})}{({\xi_{1}}{\xi_{2}}
(k_{1}+k_{2})^{2}+{\xi_{1}}{\xi_{3}}k^{2}_{1}+{\xi_{2}}{\xi_{3}}k^{2}_{2}-m^{2})}
\end{eqnarray}\\

Then due to the transversality conditions for $Z$-boson and real photons
$(k_{1}+k_{2})_{\lambda}\epsilon_{\lambda}=0$,
$\epsilon_{1\mu}k_{1\mu}=0$, $\epsilon_{2\nu}k_{2\nu}=0$ and to the
conditions for the real photons $k^{2}_{1}=0$, $k^{2}_{2}=0$ we get
the Landau theorem result $S_{Z\gamma\gamma}=0$. But in our case one
of the photons (e.g. with index 2) is virtual, as well as
$Z$-boson. Therefore the initial $S$-matrix Eq(\ref{S1}) will give
nonzero result.

Returning to our former variables $k, q$ we see that the first
term in Eq(\ref{S3}) is proportional to
\begin{equation}\label{V1}
V_{0}(q-k)(q-k)_{0}\sim{(q_{0}-k_{0})\delta(q_{0}-k_{0})=0}
\end{equation}
and is therefore absent.

Thus $S_{\mu\nu\lambda}(k_{1},k_{2})$ in our case reduces to:
\begin{equation}\label{S5}
S_{\mu\nu\lambda}(k,q)=-J_{011}(k,q)
({\varepsilon_{\lambda\mu\alpha\beta}}k_{\alpha}q_{\beta}q_{\nu}+q^{2}{\varepsilon_{\lambda\mu\nu\alpha}}k_{\alpha})
\end{equation}

Integrating over the time variables in Eq(\ref{S1}), reducing to
the three-dimensional vectors and using the three-dimensional notations
$\gamma_{0}\vec{\gamma}=\vec{\alpha}$,
$\varepsilon_{0\mu\nu\tau}=-\varepsilon_{\mu\nu\tau}$
($\mu, \nu, \tau=1,2,3$) results in
\begin{widetext}
\begin{eqnarray}\label{S6}
S=-4\pi{e}^{3}A_{PNC}\delta(E_{f}-E_{in}-\omega_{0})\sqrt{\frac{4\pi}{2\omega_{0}}}
\int{\frac{d^{3}p_{1}'}{(2\pi)^{3}}}
\frac{d^{3}p_{1}}{(2\pi)^{3}} \hspace{1mm}
\Psi^{+}_{n's}(\vec{p}_{1}) J_{011}(\vec{k},\vec{q})
\left[\frac{(\vec{\epsilon}\cdot[\vec{k}\times{\vec{q}}])(\vec{\alpha}\cdot\vec{q})}{q^{2}}+
(\vec{\epsilon}\cdot[\vec{\alpha}{\times{\vec{k}}}])\right]\Psi_{ns}(\vec{p}_{1}')
\end{eqnarray}
\end{widetext}

In the following we represent the $S$-matrix element in the
nonrelativistic limit which is obviously justified in case of
Cs atom. Recalling that the lower component $\chi$ of the Dirac
wave function could be expressed via the upper one as
$\chi=\frac{(\vec{\sigma}\cdot\vec{p})}{2m}\varphi$ and using
properties of Pauli-matrices
$(\vec{\sigma}\cdot\vec{a})(\vec{\sigma}\cdot\vec{b})=(\vec{a}\cdot\vec{b})+i(\vec{\sigma}\cdot{[\vec{a}\times{\vec{b}}]})$
we obtain the following expression for the square bracket in
Eq.(\ref{S6}) (without the factor $1/2m_{e}$)

\begin{equation}\label{P1}
(\vec{q}\cdot\vec{P})
\frac{(\vec{q}\cdot[\vec{\epsilon}\times{\vec{k}}])}{\vec{q}^{2}}
+(\vec{P}\cdot[\vec{k}\times{\vec{\epsilon}}])
+i(\vec{\sigma}\cdot\vec{k})(\vec{q}\cdot\vec{\epsilon})-
i(\vec{\sigma}\cdot\vec{\epsilon})(\vec{q}\cdot\vec{k})
\end{equation}
where $\vec{P}\equiv{\vec{p}_{1}+\vec{p}'_{1}}$ and
$\vec{q}=\vec{p}_{1}-\vec{p}'_{1}$. Expression (\ref{P1})
changes sign under the inversion
 $\vec{p}_{1}\rightarrow{-\vec{p}_{1}};
\vec{p}'_{1}\rightarrow{-\vec{p}'_{1}}$.

Then, remembering that the wave functions $\Psi_{n's}, \Psi_{ns}$
are of the same parity, the only reason for the whole expression
(\ref{S6}) not to be zero is the presence of the scalar product
$(\vec{k}_{1}\cdot\vec{k}_{2})=-(\vec{k}\cdot\vec{q})$ in the denominator of Eq(\ref{Int1}).
In the three-dimensional notations we have to analyze the integral:
\begin{eqnarray}\label{Int3}
  I(\vec{k},\vec{q})=-\frac{1}{\pi^{2}}\int^{1}_{0}d\xi_{1}\int^{1-\xi_{1}}_{0}d\xi_{2} \\\nonumber
  \times\frac{\xi_{1}(\xi_{1}+\xi_{2}-1)}{-m_{e}^{2}-2\xi_{1}\xi_{2}(\vec{k}\cdot\vec{q})+\xi_{1}(1-\xi_{1})\vec{q}^{2}}
\end{eqnarray}

The order of magnitude for the atomic electron momenta is
$|\vec{q}|\sim{m_{e}\alpha{Z}}$ and the order of magnitude for the
emitted photon momenta is $|k|=\omega\sim{m_{e}(\alpha{Z})^{2}}$.
Therefore, in the nonrelativistic  limit we have to expand the denominator in
Eq. (\ref{Int3}) and keep only the leading nonvanishing term.

This yields
\begin{widetext}
\begin{eqnarray}\label{S7}
S=-4\pi{e}^{3}A_{PNC}\delta(E_{f}-E_{in}-\omega_{0})\sqrt{\frac{4\pi}{2\omega_{0}}}
\frac{I}{m_{e}^{5}}
 \int{\frac{d^{3}p_{1}'}{(2\pi)^{3}}}
 \frac{d^{3}p_{1}}{(2\pi)^{3}}
\hspace{1mm} \Psi^{+}_{n's}(\vec{p}_{1})
[-i(\vec{\sigma}\cdot\vec{\epsilon})(\vec{q}\cdot\vec{k})^{2}]\Psi_{ns}(\vec{p}_{1}')
\end{eqnarray}
\end{widetext}
where
$
I=-1/\pi^{2}\int^{1}_{0}d\xi_{1}
\int^{1-\xi_{1}}_{0}d\xi_{2}(\xi^{2}_{1}\xi_{2})
(\xi_{1}+\xi_{2}-1)=1/{360\pi^{2}}
$

\newpage After the Fourier-transform to the coordinate representation in
Eq(\ref{S7}) and using the standard relation between $S$-matrix and the PNC-amplitude $E_{PNC}$\\ $S=-2\pi{i}E_{PNC}\delta(E_{n's}-E_{ns}-\omega)$, we
get following expression for $E_{PNC}$:

\begin{widetext}
\begin{eqnarray}\label{E}
E_{PNC}=-2i{e}^{3}\frac{(G_{F}m_{p}^{2})Q_{W}}{2\sqrt{2}}(m_{e}/m_{p})^{2}\frac{\sqrt{2\pi}}{360\pi^{2}}\frac{\omega^{3/2}_{0}}{m_{e}}
(\vec{\sigma}\cdot\vec{\epsilon})\varphi^{*}_{6s}(0)\varphi^{''}_{7s}(0)
\end{eqnarray}
\end{widetext}

We would like to note that diagram in Fig. 1.c corresponds to the
transition of the bound electron  between states of the same
parity $n's, ns$, in particular between $7s$ and $6s$ states in Cs atom,
like in experiments \cite{B, W}. But the resulting $S$-matrix
element in Eq. (\ref{E}) is proportional to $(\vec{\sigma}\cdot\vec{\epsilon})$, i.e. the
magnetic dipole moment of the electron
$\hat{\mu}=\frac12\mu_{0}\hat{\sigma}$ emits an electric type of
photon $\vec{\epsilon}$. This could be considered as a unique effect
which occurs due to the parity violation in atoms. Note, that no other radiative correction to the atomic PNC effect, evaluated up to now, can be interpreted in such a way. The $T$-invariance of
$E_{PNC}$ is satisfied due to the presence of the imaginary unit
in Eq. (\ref{E}).

For the probability of the process, combining two amplitudes [Figs 1.a and 1.c correspondingly],
we get
\begin{equation}\label{prob1}
W_{7s\rightarrow{6s}}=W_{M1}+\frac{1}{2j_{0}+1}\sum_{m_{0}m_{1}}2Re\left[E_{M1}E_{PNC}\right]+O\left(E^{2}_{PNC}\right)
\end{equation}
where $m_{0}, m_{1}$ are the angular momentum projections for the
initial and the final electron states.

Performing the summation over the electron angular momentum projections
$m_{0}, m_{1}$ and applying  the Wigner-Eckart
theorem to the product\\
$E_{M1}E_{PNC}\sim{<n's|\vec{\mu}(\vec{\nu}\times{\vec{\epsilon}})|ns>^{*}<n's|(\vec{\sigma}\cdot{\vec{\epsilon}})|ns>}$
we get the final answer in the form
\begin{equation}\label{prob2}
W_{7s\rightarrow{6s}}=W_{M1}(1+R(\vec{\nu}\cdot{\vec{s}_{ph}}))
\end{equation}
where $\nu=\vec{k}/|\vec{k}|$,
$\vec{s}_{ph}=i[\vec{\epsilon}{\times{\vec{\epsilon}^{*}}}]$ is
the spin of the photon and $R$ is so called ''degree of the parity
violation''. In our case $R$ is equal to the ratio $E_{PNC}/E_{M1}$, where the amplitudes are expressed via the angular reduced matrix elements.

Using the estimate $\varphi(0)\varphi''(0)\sim{\alpha}^{5}Z^{3}$ for neutral atoms
and $\varphi(0)\varphi''(0)\sim{\alpha}^{5}Z^{5}$ for HCI \cite{Landau2} we get
following result for the anomaly contribution to the
PNC-amplitudes in the neutral atoms (Fig. 1.c):

\begin{equation}\label{eval1}
E^{A_{atoms}}_{PNC}\sim{\frac{1}{360\pi^{2}}\left(\frac{m_{e}}{m_{p}}\right)^{2}\alpha^{3/2}(G_{F}m_{p}^{2})Q_{W}\alpha^{5}Z^{3}}
\end{equation}
and the estimate
\begin{equation}\label{eval2}
E^{A_{HCI}}_{PNC}\sim{\frac{1}{360\pi^{2}}\left(\frac{m_{e}}{m_{p}}\right)^{2}\alpha^{3/2}(G_{F}m_{p}^{2})Q_{W}\alpha^{5}Z^{5}}
\end{equation}
for the anomaly contribution to the PNC effects in HCI.

Using a well-known estimate for the PNC-amplitude \cite{X} (Fig.
1.b) in neutral atoms
\begin{equation}\label{eval3}
E^{B_{atoms}}_{PNC}\sim{\left(\frac{m_{e}}{m_{p}}\right)^{2}\alpha^{3/2}Z^{2}(G_{F}m_{p}^{2})Q_{W}}
\end{equation}
we get for the relative AA contribution a negligible value
$\sim(10)^{-3}\alpha^{5}Z$.
In the $H$-like HCI this relative contribution will be on the order of
$\sim(10)^{-3}\alpha$.

In conclusion we should stress that the observation of the AA
contribution to the PNC effects would be of a special interest
since it would be an observation of AA in atomic physics.

The authors are grateful to the participants of the Petersburg Nuclear Physics Institute Theoretical Seminar Ya. Asimov, D. Diakonov,
E. Drukarev, I. Dyatlov, L. Lipatov and N. Ural'tsev for many helpful remarks; they are also indebted to G. Pl\"{u}nien for the valuable discussions and for the hospitality during their stay in TU Dresden.

This work was supported by RFBR Grant No.
08-02-00026. G.~S. was supported also by non-profit foundation
''Dynasty'' and L.~L. was supported by the PDSPHS-MES of RF Grant No. 2.1.1/1136.

\newpage

\end{document}